# Maximum Entropy Method for Solving the Turbulent Channel Flow Problem

T.-W. Lee*

*Mechanical and Aerospace Engineering, SEMTE, Arizona State University, Tempe, AZ, 85287*

**Abstract-** There are two components in this work that allow solutions of the turbulent channel problem: one is the Galilean-transformed Navier-Stokes equation which gives a theoretical expression for the Reynolds stress (u'v'); and the second the maximum entropy principle which provides the spatial distribution of turbulent kinetic energy. The first concept transforms the momentum balance for a control volume moving at the local mean velocity, breaking the momentum exchange down to its basic components, u'v', u'$^2$, pressure and viscous forces. The Reynolds stress gradient budget confirms this alternative interpretation of the turbulence momentum balance, as validated with DNS data. The second concept of maximum entropy principle states that turbulent kinetic energy in fully-developed flows will distribute itself until the maximum entropy is attained while conforming to the physical constraints. By equating the maximum entropy state with maximum allowable (viscous) dissipation at a given Reynolds number, along with other constraints, we arrive at function forms (inner and outer) for the turbulent kinetic energy. This allows us to compute the Reynolds stress, then integrate it to obtain the velocity profiles in channel flows. The results agree well with direct numerical simulation (DNS) data at $Re_\tau$ = 400 and 1000.

*T.-W. Lee
Mechanical and Aerospace Engineering, SEMTE
Arizona State University
Tempe, AZ 85287-6106
Email: attwl@asu.edu

**INTRODUCTION**

Analytical solutions to turbulence problems have become a rarified genre, in part due to rapid advances in numerics that can solve many problems of fundamental and practical significance. We have taken an alternate route for solving turbulence problems with some modest success, in deriving the turbulence energy spectra from the maximum entropy principle [1] and in determining the Reynolds stress from the first principles [2-4]. In this work, we present a related unorthodox, but functional, method for solving the turbulence channel flow problem. The starting point is the Galilean-transformed Navier-Stokes equations [2]. To illustrate, for simple boundary layer flows, we have:

$$\frac{\partial(u^2)}{\partial x} + \frac{\partial(uv)}{\partial y} = -\frac{1}{\rho}\frac{dp}{dx} + \frac{1}{\nu}\frac{\partial^2 u}{\partial y^2} \qquad (1)$$

The instantaneous velocities, u and v, are typically decomposed into the time mean (U, V) and fluctuating (u', v') components, u = U + u', and v = V + v', which leads to cross-products of u' and v' (the Reynolds stress). A simplification in the Reynolds-averaged Navier-Stokes equation occurs when Galilean transform, U+u' → u' and V+v' → v', is applied. Under this transform, Eq. 1 gives the Reynolds stress (u'v').

$$\frac{d(u'v')}{dy} = -C_1 U \left[\frac{d(u'^2)}{dy} + \frac{1}{\rho}\frac{d|P|}{dy}\right] + \frac{1}{\nu}\frac{d^2 u'}{dy^2} \qquad (2)$$

If one wishes to solve for the diagonal component, $u'^2$, then we have:

$$\frac{d(u'^2)}{dy} = \frac{\left[-\frac{d(u'v')}{dy} + \frac{1}{\nu}\frac{d^2 u'}{dy^2}\right]}{C_1 U} - \frac{1}{\rho}\frac{d|P|}{dy} \tag{3}$$

In Eqs. 2 and 3, the variables can be Reynolds-averaged, except that u' is interpreted as u'$_{rms}$. Any gradient in the fluctuating velocity can cause viscous shear force in the mean, and u'$_{rms}$ is a representation of this momentum distribution. Also, d/dx has been replaced with $C_1 U d/dy$ to account for the displacement effect. This concept of converting d/dx was derived from consideration of control volume moving at the mean velocity in boundary layer flows with displacement effects [2-4], but it also works for channel flows as well (see Appendix).

Using Eq. 2, Reynolds stress can be directly computed using root fluid dynamic variables, U, u'$^2$ and P as shown in Figure 1(a), where the Reynolds stress gradient budget is plotted using the DNS data of Graham et al. [5] at Re$_\tau$ = 1000. The Reynolds stress gradient can then be integrated for u'v' and the mean velocity, which yields von Karman constants very close to the accepted value of 4.56 [4]. Conversely, the u'$^2$ gradient can be calculated as a function of the remaining variables, u'v', U and P, from Eq. 3, as shown in Figure 1(b). The spiked shape of the u'$^2$ profile, or its sharp gradient near the wall, is correctly tracked by Eq. 3. Figures 1(a) and (b) show that the Reynolds stress tensor can be expressed in terms of root turbulence variables which are related to one another through a relatively simple momentum balance (Eqs. 2 and 3). We just need sufficient number of equations or information to solve for the Reynolds stress tensor. In addition, Eqs. 2 and 3 and the momentum terms plotted in Figures 1(a) and (b) reveal the exchange of momentum where the u'$^2$ and u'v' are the principal carrier of u' momentum, one in the streamwise and the other cross-stream, respectively. The force terms, pressure and viscous, modify this primary momentum exchange. Eqs. 2 and 3 allow for physics-based "modeling" of turbulent flows; however, we can do a little better and solve for the turbulent channel flows if we had the turbulent kinetic energy, $u'^2$ and $v'^2$.

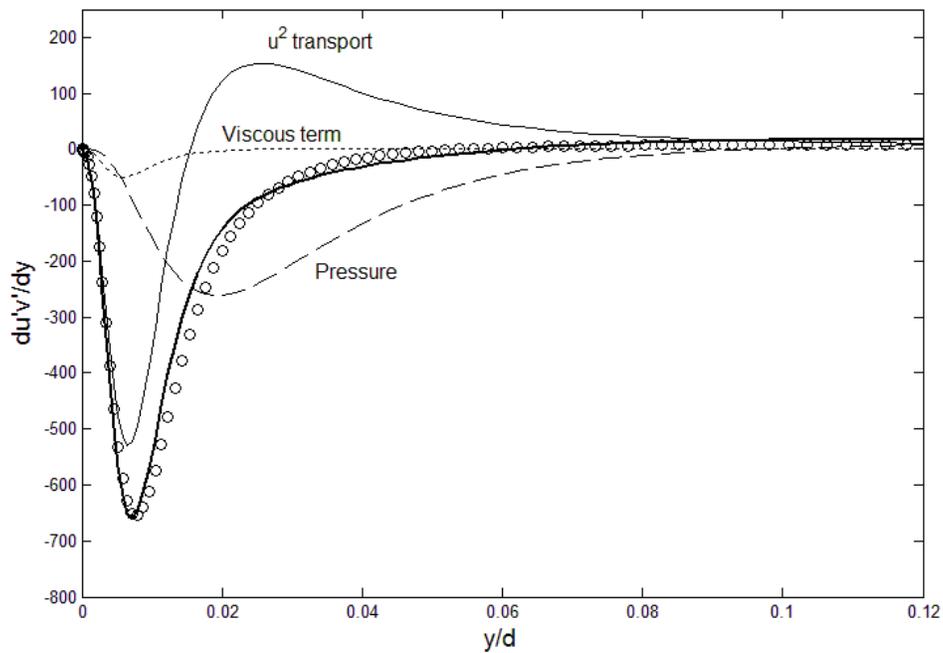

**Figure 1(a). Reynolds stress gradient budget. DNS channel flow data (circle symbol) for $Re_\tau = 1000$ [5] are used. Bold line is the RHS side of Eq. 2, with $u^2$-transport, pressure and the viscous terms combined.**

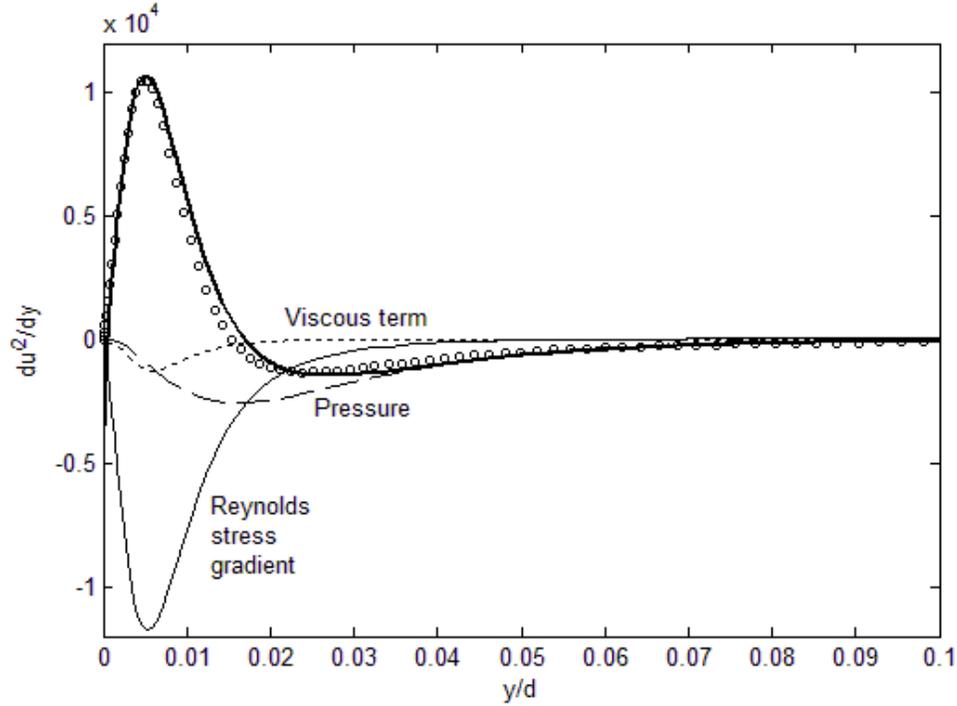

**Figure 1(b).** $u'^2$ profile obtained from Eq. 3. Reynolds stress gradient budget. DNS channel flow data (circle symbol) for $Re_\tau = 1000$ [5] are used. Bold line is the RHS side of Eq. 3, with u'v'-transport, pressure and the viscous terms combined.

Eq. 2 is an expression that relates the off-diagonal Reynolds stress term, u'v', with other turbulence variables, but for closure we still need the diagonal components, $u'^2$ and $v'^2$. For channel flows, $P = -\rho v'^2$, thus $v'^2$ is used for the pressure gradient term in Eq. 2. From the spiked shape of $u'^2$ profiles at high Reynolds numbers, both in channel and boundary layer flows, we can anticipate that finding a direct mathematical solution of $u'^2$ will not be an easy matter. In comparison, u'v' exhibits rather benign behavior, as seen in Figure 1(a). However, we have shown that the turbulence energy spectra in wavenumber space can be derived and deduced from the maximum entropy principle [1], and here we demonstrate that spatial distribution of turbulent kinetic energy ($u'^2$, $v'^2$ and $w'^2$) can also be constructed following the same principle. These profiles can then be used in Eq. 2 for theoretical solutions for turbulence channel flows. Even though we are currently at the canonical geometry stage, extensions of the current method to more complex geometry, such as backward-facing step and swirl flows, are ongoing, and will be discussed as this work evolves.

**MAXIMUM ENTROPY PRINCIPLE AND TURBULENCE**

Turbulence can be considered as a large ensemble of energetic eddies having a spectrum of energy and length scales. Due to the size of ensemble, it will come to an equilibrium state of maximum entropy under the imposed physical constraints. For turbulence energy spectra (so-called power spectra), the energy is zero at the boundary points with asymmetrical descent. The reason for this asymmetry is the physical length scale suddenly imposed on the flow at the low wavenumber and viscous dissipation at the high end. Therefore, the mechanisms for the descent to zero energy are quite different. Using this as the starting point, we have used the maximum entropy principle to derive the full turbulent energy spectra, which have a lognormal form with $k^2$ viscous dissipation at the high wavenumbers [1]. This result agrees quite well with the experimental data over nearly the entire range of Reynolds number, length and energy scales [1]. Here, we assert that the maximum entropy principle can also be applied for determination of the spatial distribution of turbulent kinetic energy, in channel flows. In channel flows, the flow evolves to the fully-developed state, which is an equilibrium state from the entropy perspective where the flow has had time to reach the maximum entropy state under the imposed physical constraints. The maximum entropy state is identified as the state where the turbulence kinetic energy is distributed in a way to achieve the maximum viscous dissipation under the physical constraints, the logic being that it is the viscous dissipation that is the primary and sole production term for entropy in isothermal flows.

Let us consider the physical attributes of the spatial energy distribution in channel flows. First, the boundary points are u'$^2$(0) = 0, and u'$^2$(d) = u'$^2_c$, where d is the channel half-width. In addition, u'$^2$ integrated over the half-width (= E) is very close to being constant, when normalized by the friction velocity (u$_\tau$), as shown in Figure 2.

$$E = \int_0^1 u'^{+2}(y) d(\frac{y}{d}) \tag{4}$$

This is a very useful feature of the normalized variable, $u'^{+2} = u'^2/u^2_\tau$. Moreover, other important variables are scalable as a function of the Reynolds number. The dissipation (= ε) is a linearly increasing function of the Reynolds number as shown in Figure 3, again when normalized by the friction velocity and integrated over the y-direction.

$$\varepsilon = \int_0^1 \left(\frac{du'^+}{dy}\right)^2 d(\frac{y}{d}) \tag{5}$$

This is due to the fact that viscosity is the limiting factor in viscous dissipation rate, and the higher the Reynolds number the flow can accommodate more viscous dissipation. The location of the peak in $u^{+2}(y)$ also scales with Re$_\tau$ with an inverse dependence, as shown in Figure 3. A similar dependence of the peak production location on Re$_\tau$ has been observed by Noor at el. [6]. The entropy interpretation of these scaling is that the turbulence energy (represented by $u'^{+2}$ or $u'^2$) distributes itself in space so that it reaches the maximum dissipation allowable at the given Reynolds number. In order to achieve high dissipation at high Reynolds numbers it develops a sharp peak which moves closer to the wall (the smaller the distance to the wall, the higher the gradient), which is the first attainable maximum entropy state. This is a lot of, and sufficient, "information" about the nature of turbulence kinetic energy in channel flows, and the maximum entropy principle is a format to combine and synthesize the available information so that the most probable energy distribution, whether in physical or wavenumber space, can be determined. Therefore, we seek $u'^2$ distribution that are consistent and unique with the above physical constraints.

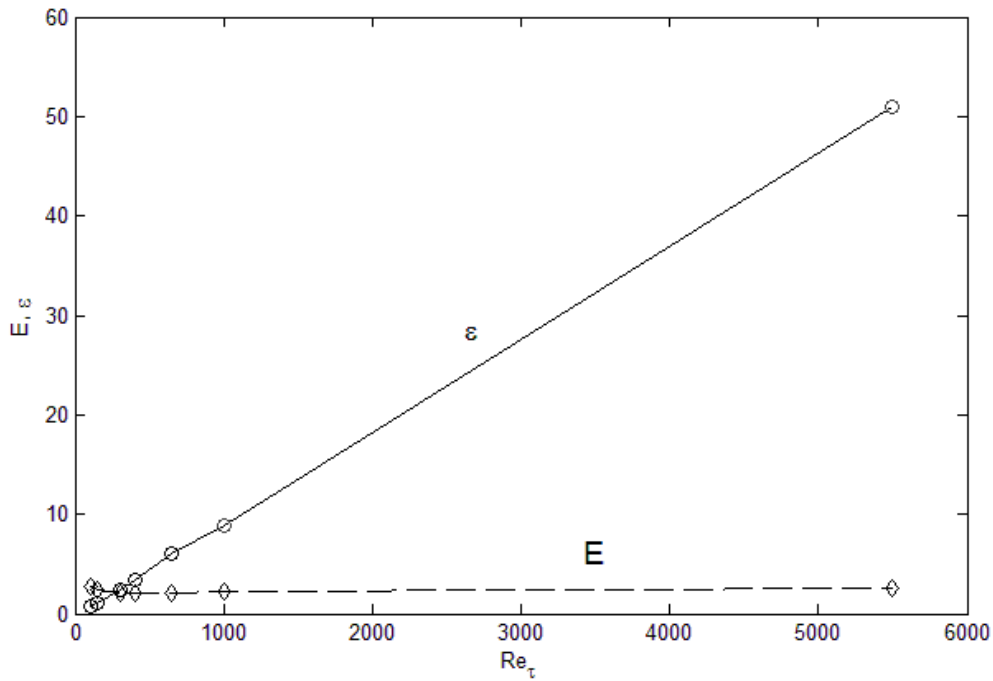

**Figure 2. Total integrated $u'^2$ (E) and dissipation ($\varepsilon$) as a function of the Reynolds numbers, from the DNS data [5, 7].**

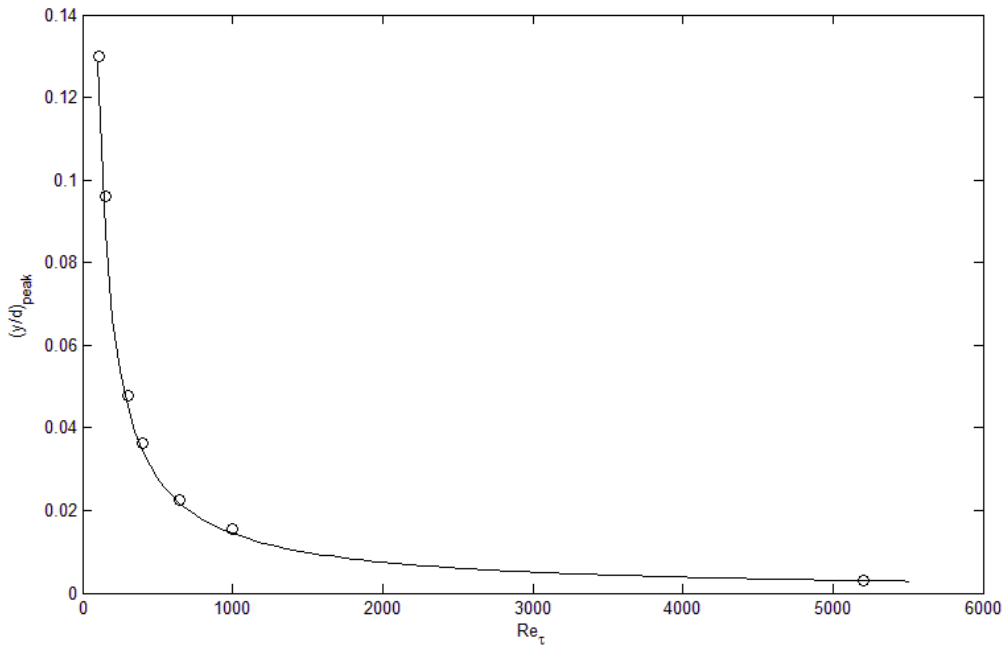

**Figure 3. The location of the u'$^2$ peak as a function of the Reynolds numbers, from the DNS data [5, 7].**

Figure 4 show such u'$^2$ profiles, constructed from the above physical constraints. For example, a combination of sharply-peaked lognormal for the inner and a beta function for the outer region works reasonably well. This is an implicit and legitimate procedure to apply the maximum energy principle [8]: select the distribution with the maximum entropy that satisfies the physical constraints. Here, the maximum entropy state is equated with that with specified dissipation, $\varepsilon$, at the given Reynolds number. The procedure is to construct the lognormal and beta functions that converges at the (y/d)$_{peak}$ location (Figure 3). Then, both E and $\varepsilon$ are computed until sufficient accuracy is achieved, relative to the data in Figure 2. Figure 4 show that this procedure is functional in reconstruction of $u'^2$ profile at a given Reynolds number for channel flows. Both the integrated energy, E, and dissipation, $\varepsilon$, are within 4% of the $Re_\tau$-dependent values from Figure 2, and the peak is located at the position specified from Figure 3. The accuracy can be improved by using a series of lognormal functions or function optimization method, which would then be continuous over the entire channel width. For boundary layer flow over flat plates, function series approach is being tested, as the $u'^2$ profiles are also sharply-peaked in such flows. For now, we use separate inner and outer functions for $u'^2$ as shown in Figure 4, in order to demonstrate the solution method. For $v'^2$ and $w'^2$ profiles, they are not subject to intense dissipation near the wall so that a single lognormal distribution complies with the constraints of zero at the wall with finite energy content and continuous decrease toward the centerline value, as shown in Figure 5.

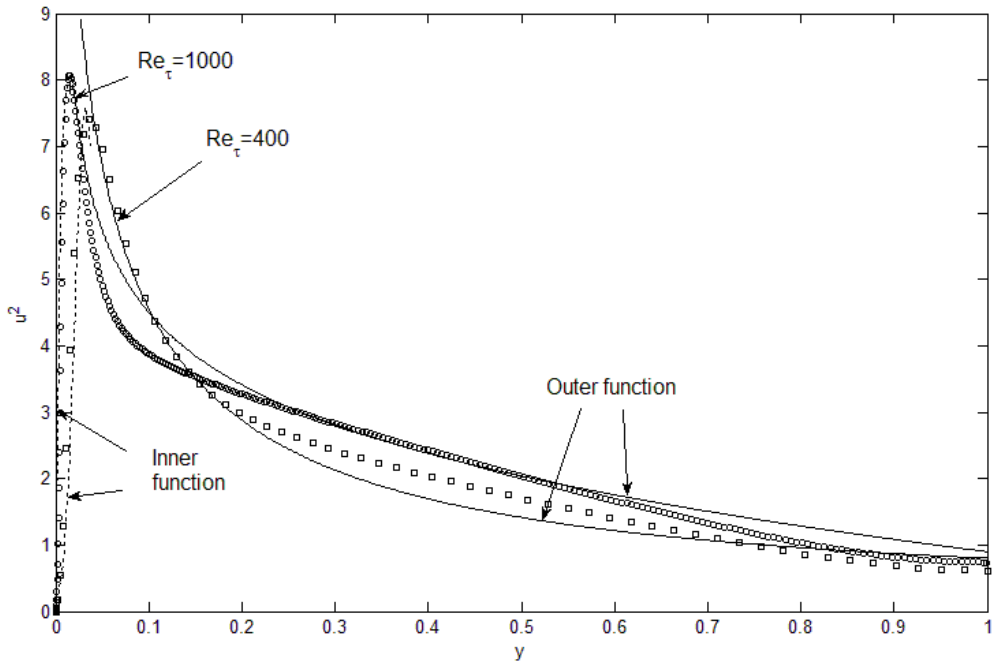

**Fig. 4. $u'^2$ profile as a combination of lognormal (inner) and beta (outer) functions. Symbols are the DNS data [5, 7].**

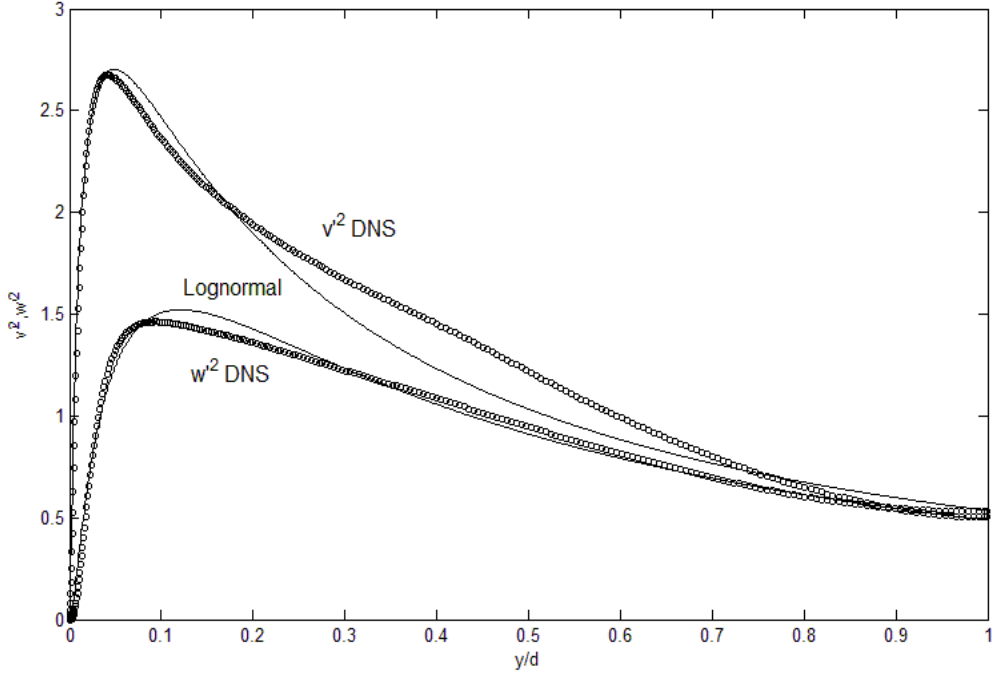

**Fig. 5. v'² and w'² profiles and lognormal functions. Symbols are the DNS data [5].**

In this way, the maximum entropy principle can be used to obtain the diagonal components of the Reynolds stress, and now we have sufficient number of equations to solve for u'v' and U through Eq. 2 and the RANS. For fully-developed channel flows, the RANS is simplified to:

Inner:  $\mu \dfrac{d^2 U}{dy^2} = \dfrac{dP}{dx} + \rho \dfrac{d(u'v')}{dy}$ (3a)

Outer:  $\mu \dfrac{dU}{dy} = \rho(u'v')$ (3b)

The solution algorithm is: we assume a reasonable (e.g. quadratic) $U(y)$ with $U(0)=0$ and $U(d)=U_c$, and insert into Eq. 2 along with $u'^2$ and $P=-\rho v'^2$ available from the maximum entropy method above. This will give us $d(u'v')/dy$, which can be integrated to $u'v'$, using Eq. 2. This is input in Eq. 3 to obtain an updated $U(y)$. This cycle is repeated until $U(y)$ converges.

The results for the Reynolds stress are shown in Figure 6. For $Re_\tau = 400$, the agreement with the DNS data is quite good. At $Re_\tau = 1000$, the Reynolds stress exhibits a rapid decrease near the wall, followed by a gradual, nearly straight, approach to zero at the centerline, which is typical of wall-bounded turbulent flows at high Reynolds numbers. The current solution deviates from this straight line, since it has been obtained from outer beta function, which has a varying slope. In addition, lognormal function is a good approximation for the $v'^2$ profile (Figure 5), but it still has a different slope in the middle part of the channel half-width. This in turn affects the pressure term in Eq. 2. Again, this is where the accuracy can be improved by finding function series that satisfies the physical constraints for $u'^2$ and $v'^2$. However, this is subject to improvements through mathematical experimentation, and not a fundamental limitation of the solution method, because sufficient constraint conditions exist to determine the turbulent kinetic energy distributions.

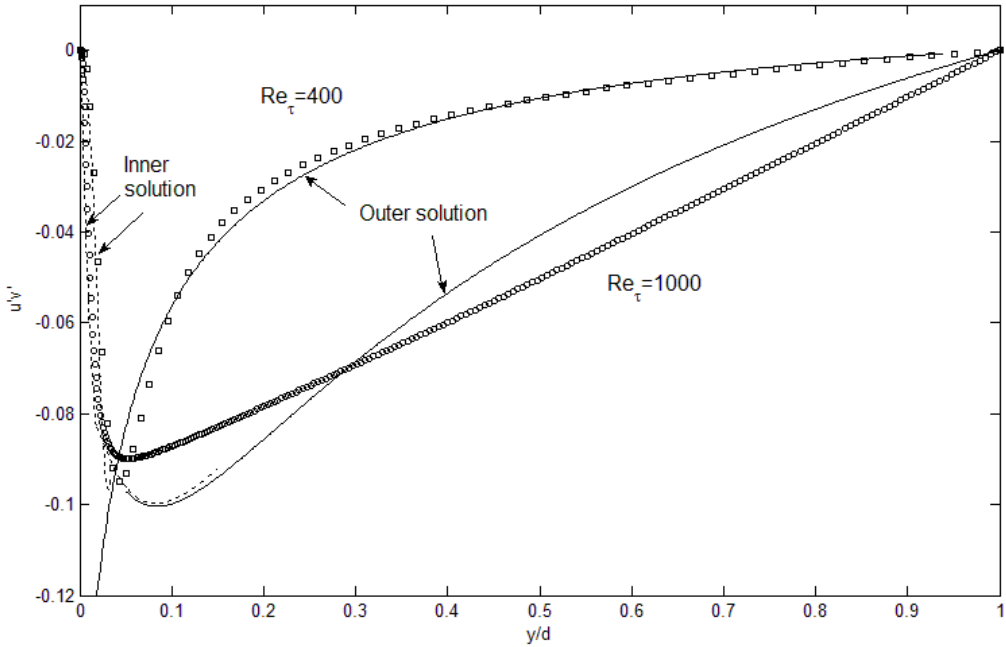

**Figure 6. The Reynolds stress profiles computed using Eq. 2, compared with DNS data [5, 7].**

For the mean velocity, the outer solutions are in very good agreement with DNS data for both $Re_\tau = 400$ and 1000, as shown in Figure 7. In the outer region, the mean velocity is essentially the integral of the Reynolds stress (times a multiplicative factor), and the integration is forgiving of minor deviations in the Reynolds stress. The mean velocity does begin to overshoot in the "overlap" region, since that is where the inner and outer functions for $u'^2$ are discontinuous. For the inner solutions, initially the solution is laminar; however, the Reynolds stress starts to exert its influence as one approaches the overlap region and the mean velocity begins to bend toward the outer solution. We can also see that the convergence between the inner and outer solutions is reasonable for $Re_\tau = 400$, leaving only a small gap where the solution is not available. For $Re_\tau = 1000$, the inaccuracy in reconstructing the $u'^2$ is propagated to $u'v'$, and then to the mean velocity, leaving a larger gap in the solution. Thus, the accuracy of the solution is obviously dependent on achieving $u'^2$ profiles that are fully compliant on the constraints discussed above.

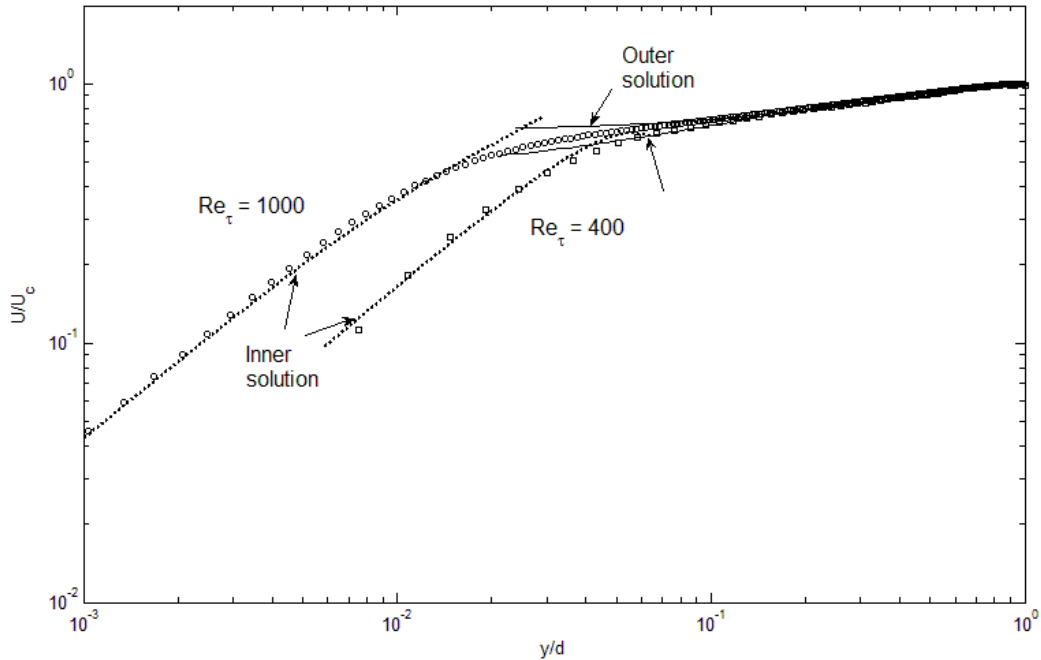

**Figure 7. The mean velocity inner and outer solutions, compared with DNS data [5, 7].**

**CONCLUSIONS**

We have used (1) the Galilean-transformed Navier-Stokes equation which gives a theoretical expression for the Reynolds stress gradient, and (2) the maximum entropy principle for the spatial distribution of turbulent kinetic energy, to obtain the inner and outer solutions to the turbulent channel problem. The Reynolds stress gradient budgets confirm the transform method, while the maximum entropy principle along with the physical constraints generate the turbulent kinetic energy profiles that are in good agreement with DNS data. This allows us to compute the Reynolds stress, which can then be integrated to obtain the velocity profiles in channel flows. The results agree well with direct numerical simulation (DNS) data at $Re_\tau = 400$ and 1000. The overlap region has not been accessed in the current inner/outer function method, but function series or function optimization can generate a single continuous function (series) which can lead to full and accurate solutions. This approach is a subject of further study in wall-bounded flows, exhibiting similar physical constraints on the turbulent energy distribution.

**REFERENCES**

[1] Lee, T.-W., Maximum entropy in turbulence, 2019, arXiv:1903.07991.

[2] T.-W., Reynolds stress in turbulent flows from a Lagrangian perspective, Journal of Physics Communications, 2018, 2, 055027.

[3] Lee, T.-W., and Park, J.E., Integral Formula for Determination of the Reynolds Stress in Canonical Flow Geometries, Progress in Turbulence VII (Eds.: Orlu, R, Talamelli, A, Oberlack, M, and Peinke, J.), pp. 147-152, 2017.

[4] Lee, T.-W., The Reynolds stress in channel flows from a Lagrangian treatment of the turbulence momentum, Proceedings of the ASME AJK-Fluids Conference, Paper 5031, July 28-Aug. 1, 2019, San Francisco, USA.

[5] Graham, J., Kanov, K, Yang, X.I.A., Lee, M.K., Malaya, N, Lalescu, C.C., Burns, R., Eyink, G, Szalay, A, Moser, R.D. Moser,Meneveau, C., A Web Services-accessible database of turbulent channel flow and its use for testing a new integral wall model for LES, Journal of Turbulence, 2016, 17(2), 181-215.

[6] Noor, A, Afzal, N., AbuSheena, S., and Bushra, A.A., Turbulent energy production peak and its location from inner most log law or power law velocity in a turbulent channel/pipe and Couette flows, European Journal of Physics, B/Fluids, 2018, Vol. 67, pp. 178-184.

[7] Iwamoto, K., Sasaki, Y., Nobuhide K., Reynolds number effects on wall turbulence: toward effective feedback control, International Journal of Heat and Fluid Flows, 2002, 23, 678-689. Also, http://thtlab.jp/DNS/dns_database.htm.

[8] Cover, T. and Thomas, J. (1991) Elements of Information Theory. John Wiley and Sons, Inc.

**APPENDIX**

The conversion, d/dx → $C_1$Ud/dy, was based on the boundary layer displacement effect for a moving control volume as shown in Figure A1. For channel flows the flow is bounded and there is no displacement of the turbulence variables as one travels in the streamwise direction. However, the Galiean transform can be performed at any line of motion, and if we choose a slightly mis-directed path (U* and v*) for the control volume as shown in Figure A2, we obtain the same transform.

For a small angle, θ << 1, v* <<U and $U* \approx U$. Then,

$$\frac{\partial}{\partial x} = \frac{1}{cos\theta}\frac{\partial}{\partial x*} \approx \frac{\partial}{\partial x*} \qquad (A1)$$

$$\frac{\partial}{\partial y} = \frac{1}{cos\theta}\frac{\partial}{\partial y*} \approx \frac{\partial}{\partial y*} \qquad (A2)$$

For variable, f, we have

$$\frac{\frac{\partial f}{\partial y*}}{\frac{\partial f}{\partial x*}} \approx \frac{\frac{\partial f}{\partial y}}{\frac{\partial f}{\partial x}} = tan\theta = \frac{v*}{U*} \approx \frac{v*}{U} \tag{A3}$$

Thus, using this offset transform, we obtain

$$\frac{\partial f}{\partial x} = \frac{U*}{v*}\frac{\partial f}{\partial y} \approx C_1 U \frac{\partial f}{\partial y} \tag{A4}$$

$C_1$ is a constant in the order and unit of v*.

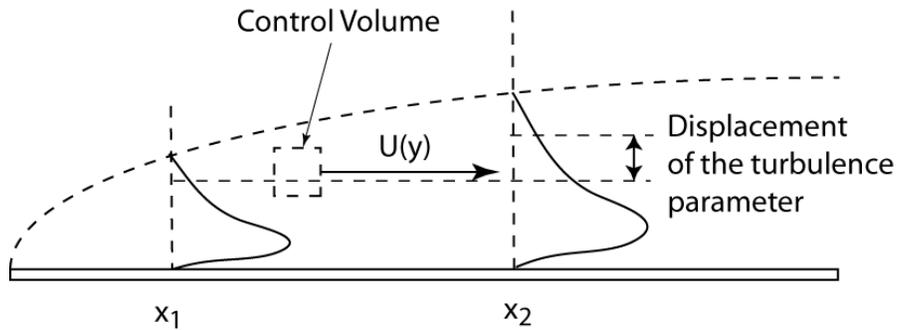

**Figure A1. Illustration of the displacement effect leading to d/dx → $C_1$Ud/dy transform.**

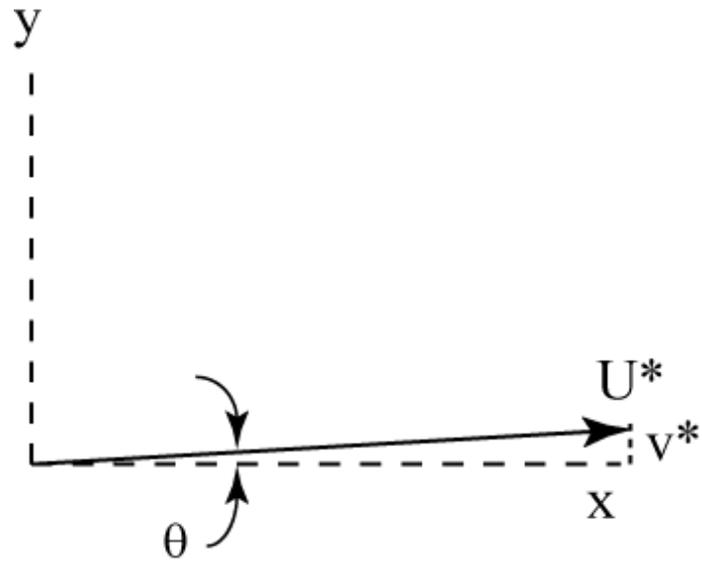

**Figure A2. Off-set line of motion for the control volume, for "probing" the d/dy gradient.**